\documentclass[twocolumn,showpacs,aps,prl,superscriptaddress,USletter]{revtex4}
\usepackage{graphicx}
\usepackage{dcolumn}
\usepackage{amsmath}
\usepackage{epsfig}

\input pubboard/babarsym.tex

\RequirePackage{xspace}

\newcommand{\calP}{\ensuremath{{\cal P}}}

\newcommand{\pvec}{{\bf p}}

\newcommand{\fL}{\ensuremath{f_L}}

\newcommand{\acp}{\ensuremath{\calA_{ch}}}

\newcommand{\calB}{\ensuremath{{\cal B}}}

\newcommand{\timesix}{\ensuremath{\times10^{6}}}
\newcommand{\timemsix}{\ensuremath{\times10^{-6}}}

\newcommand{\DE}{\ensuremath{\Delta E}}

\newcommand{\xf}{\ensuremath{{\cal F}}}
\newcommand{\hel}{\ensuremath{{\cal H}}}
\newcommand{\thetaT}{\ensuremath{\theta_{\rm T}}}
\newcommand{\costhr}{\ensuremath{\cos\thetaT}}

\newcommand\etal{{\it et al.}}
\newcommand{\half}{\ensuremath{\frac{1}{2}}}

\newcommand{\msp}{\ensuremath{\phantom{-}}}

\newcommand{\bfig}{\begin{figure}[htbpc!]}
\newcommand{\efig}{\end{figure}}
\newcommand\bef{\begin{figure}}
\newcommand\edf{\end{figure}}
\newcommand\dbline{\noalign{\vskip 0.10truecm\hrule}\noalign{\vskip 2pt}\noalign{\hrule\vskip 0.10truecm}}

\newcommand\beq{\begin{equation}}
\newcommand\eeq{\end{equation}}
\newcommand\bear{\begin{array}}
\newcommand\enar{\end{array}}
\newcommand\beqa{\begin{eqnarray}}
\newcommand\eeqa{\end{eqnarray}}
\newcommand\ben{\begin{enumerate}}
\newcommand\een{\end{enumerate}}

\newcommand{\UfourS}{\ensuremath{\Upsilon(4S)}}

\newcommand{\omtoppp}{\ensuremath{{\omega\ra\pip\pim\piz}}}

\newcommand{\Kst}{\ensuremath{K^*}}

\newcommand{\Ktwost}{\ensuremath{\Kstar_2(1430)}}
\newcommand{\Kstp}{\ensuremath{\Kstarp}}
\newcommand{\Kzerstp}{\ensuremath{\Kstar_0(1430)^+}}
\newcommand{\Ktwostp}{\ensuremath{\Kstar_2(1430)^+}}
\newcommand{\Kstz}{\ensuremath{\Kstarz}}
\newcommand{\Kzerstz}{\ensuremath{\Kstar_0(1430)^0}}
\newcommand{\Ktwostz}{\ensuremath{\Kstar_2(1430)^0}}
   \newcommand{\KstpKppiz}{\ensuremath{\Kstarp_{K^+\pi^0}}}
   \newcommand{\KstptoKppiz}{\ensuremath{\Kstarp\ra K^+\pi^0}}
   \newcommand{\KstpKspip}{\ensuremath{\Kstarp_{\KS\pi^+}}}
   \newcommand{\KstptoKspip}{\ensuremath{\Kstarp\ra \KS\pi^+}}
   \newcommand{\KstzKppim}{\ensuremath{\Kstarz_{K^+\pi^-}}}
   \newcommand{\KstztoKppim}{\ensuremath{\Kstarz\ra K^+\pi^-}}

   \newcommand{\rhop}{\ensuremath{\rho^+}}
   
\newcommand{\kzs}{\ensuremath{\KS}}

\newcommand{\omegaKtst}{\ensuremath{\B\ra\omega\Ktwost}}
\providecommand{\fomegaKstz}{\ensuremath{\omega\Kstz}}

\providecommand{\fomegaKzstz}{\ensuremath{\omega\Kzerstz}}
\providecommand{\omegaKzstz}{\ensuremath{\Bz\ra\fomegaKzstz}}
\newcommand{\BomegaKzstz}{\ensuremath{\calB(\omegaKzstz)}}
\providecommand{\fomegaKtstz}{\ensuremath{\omega\Ktwostz}}

\providecommand{\fomegaKstp}{\ensuremath{\omega\Kstp}}

\providecommand{\fomegaKzstp}{\ensuremath{\omega\Kzerstp}}
\providecommand{\omegaKzstp}{\ensuremath{\Bp\ra\fomegaKzstp}}
\newcommand{\BomegaKzstp}{\ensuremath{\calB(\omegaKzstp)}}
\providecommand{\fomegaKtstp}{\ensuremath{\omega\Ktwostp}}

\providecommand{\romegaKstp}{\ensuremath{3.5^{+2.5}_{-2.0}\pm0.7}}

\providecommand{\ulomegaKstp}{\ensuremath{7.4}}

\newcommand{\romegaKstz}{\ensuremath{3.4^{+1.8}_{-1.6}\pm0.4}}

\newcommand{\fomegaKstpKspip}{\ensuremath{\omega\Kstp_{\KS \pip}}}

\newcommand{\fomegaKtstpKspip}{\ensuremath{\omega\Ktwostp_{\KS\pip}}}

\newcommand{\fomegaKstpKppiz}{\ensuremath{\omega \Kstp_{\Kp\piz}}}

\newcommand{\fomegaKtstpKppiz}{\ensuremath{\omega \Ktwostp_{\Kp\piz}}}

\newcommand{\fomegarhop}{\ensuremath{\omega\rho^+}\xspace}
\newcommand{\omegarhop}{\ensuremath{\Bp\ra\fomegarhop}\xspace}

\newcommand{\romegarhop}{\ensuremath{13.5^{+3.8}_{-3.5}\pm1.7}\xspace}

\newcommand{\somegarhop}{\ensuremath{xx}\xspace}
\newcommand{\Aomegarhop}{\ensuremath{0.05\pm 0.26\pm 0.02}}
\newcommand{\fLomegarhop}{\ensuremath{0.88^{+0.12}_{-0.15}}}

\newcommand{\fomegarhoz}{\ensuremath{\omega\rho^0}}
\newcommand{\omegarhoz}{\ensuremath{\Bz\ra\fomegarhoz}}

\newcommand{\romegarhoz}{\ensuremath{0.59^{+1.3}_{-1.1}\pm0.35}}

\newcommand{\ulomegarhoz}{\ensuremath{3.3}\xspace}

\newcommand{\fomegafz}{\ensuremath{\omega f_0}\xspace}
\newcommand{\omegafz}{\ensuremath{\Bz\ra\fomegafz}\xspace}

\newcommand{\romegafz}{\ensuremath{0.xx^{+x.x}_{-x.x}\pm0.xx}\xspace}

\newcommand{\ulomegafz}{\ensuremath{x.x}\xspace}

\newcommand{\jhep}[1]{{\it JHEP}\ #1}

\newcommand{\helo}{\ensuremath{\hel_{\omega}}}
\newcommand{\helrho}{\ensuremath{\hel_{\pi\pi}}}

\newcommand{\KpiSwave}{\ensuremath{(K\pi)^*_0}}
\newcommand{\KpiSwavez}{\ensuremath{(K\pi)^{*0}_0}}
\newcommand{\KpiSwavep}{\ensuremath{(K\pi)^{*+}_0}}
\newcommand{\KpizSwavep}{\ensuremath{(\Kp\piz)^{*+}_0}}
\newcommand{\KspiSwavep}{\ensuremath{(\KS\pip)^{*+}_0}}
\newcommand{\fomegaKpiSwavez}{\ensuremath{\omega\KpiSwavez}}
\newcommand{\fomegaKpiSwavep}{\ensuremath{\omega\KpiSwavep}}
\newcommand{\fomegaKpiSwavepKspip}{\ensuremath{\omega\KspiSwavep}}
\newcommand{\fomegaKpiSwavepKppiz}{\ensuremath{\omega\KpizSwavep}}

\renewcommand{\romegarhop}{\ensuremath{15.9\pm1.6\pm1.4}} 
\renewcommand{\somegarhop}{\ensuremath{9.8}}		
\renewcommand{\Aomegarhop}{\ensuremath{-0.20\pm0.09\pm0.02}}  
\renewcommand{\fLomegarhop}{\ensuremath{0.90\pm0.05\pm0.03}}  

\renewcommand{\romegarhoz}{\ensuremath{0.8\pm0.5\pm0.2}} 
\renewcommand{\ulomegarhoz}{\ensuremath{1.6}}	

\renewcommand{\romegafz}{\ensuremath{1.0\pm0.3\pm0.1}} 
\renewcommand{\ulomegafz}{\ensuremath{1.5}}	

\renewcommand{\romegaKstz}{\ensuremath{2.2\pm0.6\pm0.2}} 
\newcommand{\somegaKstz}{\ensuremath{4.1}}		
\newcommand{\AomegaKstz}{\ensuremath{0.45\pm0.25\pm0.02}}  
\newcommand{\fLomegaKstz}{\ensuremath{0.72\pm0.14\pm0.02}} 

\newcommand{\romegaKpiSwavez}{\ensuremath{18.4\pm1.8\pm1.7}}
\newcommand{\romegaKzstz}{\ensuremath{16.0\pm1.6\pm1.5\pm2.6}}
\newcommand{\somegaKpiSwavez}{\ensuremath{9.8}}	

\newcommand{\RomegaKzstz}{\ensuremath{(\romegaKzstz)\times 10^{-6}}}
\newcommand{\AomegaKpiSwavez}{\ensuremath{-0.07\pm0.09\pm0.02}}

\newcommand{\romegaKtstz}{\ensuremath{10.1\pm2.0\pm1.1}}
\newcommand{\somegaKtstz}{\ensuremath{5.0}}		
\newcommand{\AomegaKtstz}{\ensuremath{-0.37\pm0.17\pm0.02}}
\newcommand{\fLomegaKtstz}{\ensuremath{0.45\pm0.12\pm0.02}} 

\renewcommand{\romegaKstp}{\ensuremath{2.4\pm1.0\pm0.2}} 
\newcommand{\somegaKstp}{\ensuremath{2.5}}		
\providecommand{\ulomegaKstp}{\ensuremath{3.8}}

\newcommand{\AomegaKstp}{\ensuremath{0.29\pm0.35\pm0.02}}	
\newcommand{\fLomegaKstp}{\ensuremath{0.41\pm0.18\pm0.05}}      

\newcommand{\romegaKpiSwavep}{\ensuremath{27.5\pm3.0\pm2.6}} 
\newcommand{\romegaKzstp}{\ensuremath{24.0\pm2.6\pm2.2\pm3.8}} 
\newcommand{\somegaKpiSwavep}{\ensuremath{9.2}}		
\newcommand{\RomegaKzstp}{\ensuremath{(\romegaKzstp)\times 10^{-6}}}
\newcommand{\AomegaKpiSwavep}{\ensuremath{-0.10\pm0.09\pm0.02}}

\newcommand{\romegaKtstp}{\ensuremath{21.5\pm3.6\pm2.4}} 
\newcommand{\somegaKtstp}{\ensuremath{6.1}}		
\newcommand{\AomegaKtstp}{\ensuremath{0.14\pm0.15\pm0.02}}
\newcommand{\fLomegaKtstp}{\ensuremath{0.56\pm0.10\pm0.04}}      

\newcommand{\BABARPubYear}    {08}
\newcommand{\BABARPubNumber}  {056}
\newcommand{\SLACPubNumber} {13510}
\newcommand{\LANLNumber} {0901.3703 [hep-ex]}

\begin{document}


\begin{flushleft}
\babar-PUB-\BABARPubYear/\BABARPubNumber\\
SLAC-PUB-\SLACPubNumber\\
arXiv:\LANLNumber
\end{flushleft}

\title{
{\large \boldmath \bf Observation of $B$ Meson Decays to $\omega\Kstar$
and Improved Measurements for $\omega\rho$ and $\omega f_0$ }
}

%
\author{B.~Aubert}
\author{Y.~Karyotakis}
\author{J.~P.~Lees}
\author{V.~Poireau}
\author{E.~Prencipe}
\author{X.~Prudent}
\author{V.~Tisserand}
\affiliation{Laboratoire d'Annecy-le-Vieux de Physique des Particules (LAPP), Université de Savoie, CNRS/IN2P3, F-74941 Annecy-Le-Vieux, France }
\author{J.~Garra~Tico}
\author{E.~Grauges}
\affiliation{Universitat de Barcelona, Facultat de Fisica, Departament ECM, E-08028 Barcelona, Spain }
\author{L.~Lopez$^{ab}$ }
\author{A.~Palano$^{ab}$ }
\author{M.~Pappagallo$^{ab}$ }
\affiliation{INFN Sezione di Bari$^{a}$; Dipartmento di Fisica, Universit\`a di Bari$^{b}$, I-70126 Bari, Italy }
% ...
\collaboration{The \babar\ Collaboration}
\noaffiliation

\date{\today}

\begin{abstract}
We present measurements of $B$ meson decays to the final states 
$\omega\Kst$, $\omega\rho$, and $\omega f_0$, where \Kstar\ indicates a
spin 0, 1, or 2 strange meson.  The data sample corresponds
to 465$\times10^6$ \BB\ pairs collected with the \babar\ detector 
at the PEP-II \epem\ collider at SLAC.  $B$ meson decays involving 
vector-scalar, vector-vector, and vector-tensor final states are analyzed;
the latter two shed new light on the polarization of these final states.
We measure the branching fractions for nine of these decays; five are
observed for the first time.  For most decays we also measure the charge 
asymmetry and, where relevant, the longitudinal polarization \fL.
\end{abstract}


\maketitle

Studies of vector-vector ($VV$) final states in $B$ decays resulted in the
surprising observation that the longitudinal polarization fraction 
\fL\ in $B\to\phi\Kstar$ decays is $\sim$0.5, not $\sim$1 \cite{phiKstorig}.
The latter value is expected from simple helicity arguments and has been 
confirmed in the tree-dominated \cite{BRYCYlilu} $B\to\rho\rho$ decays 
\cite{rhorho} and \omegarhop\ decays \cite{PrevBABAR}.
It appears that the \fL$\sim$1 expectation is correct for
tree-dominated decays but is not generally true for decays 
where $b\to s$ loop (penguin) amplitudes are dominant.

There have been numerous attempts to understand the polarization puzzle (small 
\fL) within the Standard Model (SM)~\cite{VVBSMrefs}, and many papers have 
suggested non-SM explanations~\cite{nSMetc}.  The SM picture
improved recently with the calculation of \fL\ for most charmless $VV$ decays 
\cite{BRYCYlilu} with inclusion of non-factorizable effects and penguin
annihilation amplitudes.
Improved understanding of these effects can come from branching fraction
and \fL\ measurements in decays such as $B\to\omega\Kstar$, which is related 
to $B\to\phi\Kstar$ via SU(3) symmetry~\cite{oh}.  Among these decays, there
is evidence for only $\Bz\to\omega\Kstarz$
\cite{PrevBABAR,BelleomKst}. Information on these and related charmless
$B$ decays can be used to provide constraints on the CKM
angles $\alpha$, $\beta$, and $\gamma$~\cite{VVphaserefs}.

Further information on the polarization puzzle can come from
measurements that include the tensor meson $\Kst_2(1430)$.  
A measurement of the vector-tensor ($VT$) decay $B\to\phi\Kst_2(1430)$
\cite{BABARphiK2} finds a value of \fL\ inconsistent with 0.5 (but
consistent with 1), so a measurement of the related decay 
$B\to\omega\Kst_2(1430)$ would be interesting.  
The only theoretical predictions for these modes are from generalized 
factorization calculations \cite{KLO}; the branching fraction predictions for 
the $B\to\omega\Kst_2(1430)$ decays are $\sim(1-2)\timemsix$, but there are
no predictions for \fL.  There have been a variety of measurements
for similar $B$ decays that include the scalar meson $\Kst_0(1430)$
\cite{etaKst,BABARphiK2}.  For the scalar-vector ($SV$) decays 
$B\to\omega\Kst_0(1430)$, there are recent QCD factorization calculations
\cite{CCY} that predict branching fractions of about $10^{-6}$.

We report measurements of $B$ decays to the 
final states $\omega\Kst$, $\omega\rho$, and $\omega f_0(980)$, where \Kstar\ 
includes the spin 0, 1, and 2 states, $\Kst_0(1430)$, \Kst(892), and 
$\Kst_2(1430)$, respectively.
While a complete angular analysis of the $VV$ and  $VT$ decays would
determine helicity amplitudes fully, because of the small signal samples
we measure only \fL.  Given our uniform
azimuthal acceptance, we obtain, after integration, the angular
distributions $d^2\Gamma/(d\cos{\theta_1}d\cos{\theta_2})$:
\begin{eqnarray}
f_T\sin^2\theta_1\sin^2\theta_2 &+&4f_L\cos^2\theta_1\cos^2\theta_2,
\label{eq:VVAngDist} \\
f_T\sin^2\theta_1\sin^2\theta_2\cos^2\theta_2
&+&\frac{f_L}{3}\cos^2\theta_1(3\cos^2\theta_2-1)^2\quad\quad \label{eq:VTAngDist}
\end{eqnarray}
for the $VV$ and $VT$ \cite{Datta} decays, respectively, where $f_T=1-\fL$ 
and $\theta_1$ and $\theta_2$ are
the helicity angles in the $V$ or $T$ rest frame with respect to the boost 
axis from the $B$ rest frame.  For decays with significant signals, we
also measure the direct \CP-violating, time-integrated charge asymmetry
$\acp \equiv (\Gamma^- - \Gamma^+)/(\Gamma^- + \Gamma^+)$,
where the superscript on the $\Gamma$ corresponds to the charge of the \Bpm 
meson or the charge of the kaon for \Bz decays.

The results presented here are obtained from data collected with the 
\babar\ detector~\cite{BABARNIM} at the PEP-II asymmetric-energy $e^+e^-$ 
collider located at SLAC.  An integrated
luminosity of 424~fb$^{-1}$, corresponding to 
465\timesix\ \BB\ pairs, was recorded at the \UfourS\
resonance, with \epem\ center-of-mass (CM) energy $\sqrt{s}=10.58\ \gev$.

Charged particles from the \epem\ interactions are detected, and their
momenta measured, by five layers of double-sided
silicon microstrip detectors surrounded by a 
40-layer drift chamber, both operating in the 1.5-T magnetic
field of a superconducting solenoid. We identify photons and electrons 
using a CsI(Tl) electromagnetic calorimeter (EMC).
Further charged particle identification (PID) is provided by the average energy
loss ($dE/dx$) in the tracking devices and by an internally reflecting
ring-imaging Cherenkov detector (DIRC) covering the central region.

We reconstruct $B$-daughter candidates through their decays
$\rho^0\ra\pip\pim$, $f_0(980)\ra\pip\pim$, $\rho^+\ra\pip\piz$,
\KstztoKppim, \KstptoKppiz (\KstpKppiz), \KstptoKspip
(\KstpKspip), \omtoppp, $\piz\ra\gaga$, and
$\kzs\ra\pip\pim$. Charge-conjugate decay modes are implied unless
specifically stated.  Table \ref{tab:rescuts}\ lists the requirements on the
invariant masses of these final states.  For the $\rho$,
\Kstar, and $\omega$ selections, these mass requirements
include sidebands, as the mass values are
treated as observables in the maximum-likelihood fit described below.
For \kzs\ candidates we further require the three-dimensional flight
distance from the primary vertex to be greater than three times
its uncertainty.  Daughters of $\rho$,
\Kstar, and $\omega$ candidates are rejected if their DIRC, $dE/dx$,
and EMC PID signatures are highly consistent with protons or
electrons; kaons must have a kaon signature while the pions must not.

\begin{table}[btp]
\begin{center}
\caption{
Selection requirements on the invariant masses and helicity angles
of $B$-daughter resonances.  The helicity angle is unrestricted 
unless indicated otherwise.
}
\label{tab:rescuts}
\begin{tabular}{lcc}
\dbline
State           & inv. mass (\mev)  &    helicity angle                     \\
\dbline                                        
\KstzKppim,\KstpKspip&$750 < m_{K\pi}< 1550$     &$-0.85<\cos{\theta}<1.0$\\
\KstpKppiz      & $750 < m_{K\pi}< 1550$          &$-0.80<\cos{\theta}<1.0$ \\
$\rho^0/f_0$    & $470 < m_{\pi\pi} <1070$       &$-0.80<\cos{\theta}<0.80$\\
$\rho^+$        & $470 < m_{\pi\pi} <1070$       &$-0.70<\cos{\theta}<0.80$\\
$\omega$        & $735 < m_{\pi\pi\pi} <825$     &    \\
\piz            & $120 < m_{\gamma\gamma} < 150$ &       \\
\kzs            & $488<m_{\pi\pi} <508$          &       \\
\dbline
\end{tabular}
\vspace{-5mm}
\end{center}
\end{table}

Table \ref{tab:rescuts}\ also gives the restrictions on the \Kstar and
$\rho$ helicity angle $\theta$ imposed to avoid 
regions of large combinatorial background from low-momentum particles.
To calculate $\theta$ we take the angle relative to a
specified axis: for $\omega$, the normal to the decay plane; for $\rho$,
the positively-charged daughter momentum;
and for \Kstar, the daughter kaon momentum.

A $B$-meson candidate is characterized kinematically by the
energy-substituted mass $\mes\equiv\sqrt{(\half
s+\pvec_0\cdot\pvec_B)^2/E_0^{*2}-\pvec_B^2}$ and the energy
difference $\DE \equiv E_B^*-\half\sqrt{s}$, 
where $(E_0,\pvec_0)$ and $(E_B,\pvec_B)$ are four-momenta of
the \epem\ CM and the $B$ candidate, respectively, $s$ is the square of
the CM energy, and the asterisk denotes the \epem\ CM frame.
Signal events peak at zero for \DE, and at the $B$ mass \cite{PDG2008} 
for \mes, with a resolution for \DE\ (\mes) of 30--45 MeV ($3.0\ \mev$).
We require $|\DE|\le0.2$ GeV and $5.25\le\mes<5.29\ \gev$.

The angle \thetaT\ between the thrust axis of the $B$ candidate in the \epem\
CM frame and that of the charged tracks and neutral clusters in the rest
of the event is used to reject the dominant continuum $\epem\ra\qqbar$
($q=u,d,s,c$) background events.  The distribution of $|\costhr|$ is
sharply peaked near $1.0$ for combinations drawn from jet-like \qqbar
pairs, and is nearly uniform for the almost isotropic $B$-meson decays.
We reduce the sample sizes to 30000--65000 events by requiring
$|\costhr|<0.7$ for the $\omega\rho/f_0$ modes and $|\costhr|<0.8$ for the 
$\omega\Kstar$ modes.  Further discrimination from continuum is obtained 
with a Fisher discriminant \xf\ that combines four variables:
the polar angles, with respect to the beam axis in the \epem\ CM frame, of the 
$B$ candidate momentum and of the $B$ thrust axis; and the zeroth and second
angular moments $L_{0,2}$ of the energy flow, excluding the $B$
candidate, about the $B$ thrust axis.  The mean of \xf\ is adjusted so
that it is independent of the $B$-flavor tagging category \cite{ccbarK0}.
The moments are defined by $ L_j = \sum_i
p_i\times\left|\cos\theta_i\right|^j,$ where $\theta_i$ is the angle
with respect to the $B$ thrust axis of track or neutral cluster $i$ and
$p_i$ is its momentum.
The average number of $B$ candidates found per selected event in data is in
the range 1.1 to 1.3, depending on the final state.  We choose the
candidate with the highest value of the probability for the $B$ vertex fit.

We obtain yields and values of \fL\ and \acp\ from extended unbinned 
maximum-likelihood (ML) fits with input observables \DE, \mes, \xf, and, for the
scalar, vector or tensor meson, the invariant mass and $\hel=\cos\theta$.  
For each event $i$ and hypothesis $j$ (signal, \qqbar\ background, 
\BB\ background), we define the probability density function (PDF) with
resulting likelihood $\cal L$:
\begin{eqnarray}
\calP^i_j &=& \calP_j(\mes^i) \calP_j(\DE^i) \calP_j(\xf^i) 
 \calP_j(m_1^i,m_2^i,\hel^i_1,\hel^i_2)\,, \quad\quad \label{eq:evtL}\\
    {\cal L} &=& \frac{e^{-(\sum_j Y_j)}}{N!} \prod_{i=1}^N
\sum_j Y_j \calP_j^i\ , \label{eq:totalL}
\end{eqnarray}
where $Y_j$ is the yield of events of hypothesis $j$,
$N$ is the number of events in the sample, and the subscript 1 (2)
represents $3\pi$ ($K\pi$ or $\pi\pi$).
There are as many as three signal categories and the PDFs for each
are split into two components: correctly reconstructed
events and those where candidate particles are exchanged with
a particle from the rest of the event.  The latter component is called
self crossfeed (SXF) and its fractions are fixed to the values found in 
Monte Carlo (MC), (15--35)\%.  We find correlations among the 
observables to be small for \qqbar\ background.  
  
\begin{table*}[!bth]
\caption{
Signal yield $Y$ and its statistical uncertainty, bias $Y_0$,
detection efficiency $\epsilon$, daughter branching fraction product
$\prod\calB_i$, significance $S$ (with systematic uncertainties included), 
measured branching fraction \calB\ with statistical and systematic
errors, 90\% C.L. upper limit (U.L.), measured
or assumed \fL, and \acp.  In the case of \fomegafz, 
the quoted branching fraction is a product with $\calB(f_0\to\pi\pi)$,
which is not well known. \KpiSwave\ refers to the S-wave $K\pi$ system.
}
\label{tab:results}
\begin{tabular}{lcrrrrcccc}
\dbline
Mode            & $Y$      &$Y_0~~~~$   &$\epsilon~~~$ &$\prod\calB_i$ & $S~$       &  \calB        & \calB\ U.L. &  ~~$f_L$  & \acp \\
                & (events) & (events)   &(\%)$~$     & (\%)$~$          &$(\sigma)$ & $(10^{-6})$   & $(10^{-6})$ &   & \\
\dbline
\fomegaKstz    &$101\pm25$ &$8\pm4~$&15.2 &59.5 &\somegaKstz &\romegaKstz &$-$          &\fLomegaKstz &~~\msp\AomegaKstz \\
\fomegaKstp    &           &       &     &     &\somegaKstp &\romegaKstp &\ulomegaKstp &\fLomegaKstp &~~\msp\AomegaKstp \\
~~\fomegaKstpKppiz&$72\pm24$&$3\pm2~$ &10.4 &29.7 &3.7 &$4.8\pm1.7$ &    &$0.37\pm0.18$ &$\msp0.22\pm0.33$ \\
~~\fomegaKstpKspip&$8\pm16$ &$0\pm1~$ &13.6 &20.6 &0.5 &$0.6\pm1.2$ &    &0.5 fixed     &$-$            \\
\fomegaKpiSwavez  &$540\pm47$ &$49\pm25$& 9.7 &59.5 &\somegaKpiSwavez &\romegaKpiSwavez &$-$&$-$&~~\AomegaKpiSwavez \\
\fomegaKpiSwavep    &           &        &     &     &\somegaKpiSwavep &\romegaKpiSwavep &$-$&$-$&~~\AomegaKpiSwavep \\
~~\fomegaKpiSwavepKppiz&$191\pm36$&$18\pm9$&6.4&29.7 &5.9 &$19.6\pm4.1$&    &$-$ &$-0.38\pm0.19$ \\
~~\fomegaKpiSwavepKspip&$357\pm39$&$34\pm17$&9.1&20.6 &10.6&$37.1\pm4.5$&    &$-$ &$-0.01\pm0.10$            \\
\fomegaKtstz    &$185\pm32$ &$19\pm10$&11.9 &29.7 &\somegaKtstz &\romegaKtstz &$-$           &\fLomegaKtstz &~~\AomegaKtstz \\
\fomegaKtstp    &           &   &     &     &\somegaKtstp&\romegaKtstp &$-$           &\fLomegaKtstp &~~\msp\AomegaKtstp \\
~~\fomegaKtstpKppiz&$182\pm30$&$6\pm3~$ &8.2&14.9 &7.2 &$31.0\pm5.2$ &    &$0.52\pm0.10$ &$\msp0.17\pm0.16$ \\
~~\fomegaKtstpKspip&$64\pm25$ &$10\pm5~$ &10.1&10.3 &2.4 &$11.2\pm4.9$ &    &$0.76\pm0.26$ &$-0.04\pm0.35$  \\
\fomegarhoz &$30^{+21}_{-18}$&$-3\pm2~$&9.5 &89.2 &1.9 &\romegarhoz &\ulomegarhoz &0.8 fixed      & --- \\
\fomegafz   &$37^{+14}_{-12}$&$1\pm1~$&14.4 &59.5 &4.5 &\romegafz   &\ulomegafz& ---       & --- \\
\fomegarhop &$411\pm43$      &$27\pm14$&5.8 &89.2 &\somegarhop &\romegarhop & --- & \fLomegarhop  & ~~\Aomegarhop \\
\dbline
\end{tabular}
\vspace{-5mm}
\end{table*}

From MC simulation \cite{geant} we form a sample of the most
relevant charmless \BB\ backgrounds (20--35 modes for each signal
final state).  We include a fixed yield (70--200 events, derived from 
MC with known or estimated branching fractions) for these in the fit described 
below.  For \omegarhop\ we also introduce a component for nonresonant
$\omega\pip\piz$ background; for the other decays nonresonant backgrounds
are smaller and are
included in the charmless \BB\ sample.  The magnitude of the nonresonant 
component is fixed in each fit as determined from fits to regions of
higher $\pi\pi$ or $K\pi$ mass.  For the $\omega\rho$ modes, we also include a
sample of $b\to c$ backgrounds; for the other modes, this component
is not used since it is not clearly distinguishable from \qqbar background.

Signal is also simulated with MC; for the \KpiSwave\ line shape, we use a LASS 
model~\cite{LASS,Latham} which consists of the $\Kst_0(1430)$ 
resonance together with an effective-range nonresonant component.
For the $f_0(980)$, we use a Breit-Wigner shape with parameters taken
from Ref.~\cite{E791fz}.

The PDF for resonances in the signal takes the form
$\calP_{1,{\rm sig}}(m_1^i)\calP_{2,{\rm sig}}(m_2^i){\cal
Q}(\hel^i_1,\hel^i_2)$ with ${\cal Q}$ given by Eq.~\ref{eq:VVAngDist} 
or \ref{eq:VTAngDist}, modified to account for detector acceptance.  For 
\qqbar\ background we use for each resonance independently
$\calP_{\qqbar}(m_k^i,\hel_k^i)=\calP_{\qqbar}(m_k^i)\calP_{\qqbar}(\hel_k^i)$,
where $\calP_{\qqbar}(m_k^i)$ is a sum of true resonance and
combinatorial mass terms.  The PDFs for \BB\ background have a similar form.

For the signal, \BB\ background, and nonresonant background 
components we determine the PDF parameters from simulation.  We study large 
data control samples of $\Bp\to\Dzb\pip$ and $\Bp\to\Dzb\rhop$ decays with 
$\Dzb\to\Kp\pim\piz$ to check the simulated resolutions in \DE\ and \mes,
and adjust the PDF parameters to account for small differences.  For the 
continuum background we use (\mes,\,\DE) sideband data to obtain initial 
values of the parameters, and leave them free to vary in the ML fit.

The parameters that are allowed to vary in the fit include the signal
and \qqbar background yields, \fL\ (for all $VV$ and $VT$ modes except 
\omegarhoz), continuum background PDF parameters, and, for $\omega\rho$, 
the $b\to c$ background yield.  Since there is not a significant yield
for \omegarhoz, we fix \fL\ to a value that is consistent with \textit{a
priori} expectations \cite{BRYCYlilu} (see Table \ref{tab:results}).
For all modes except \omegarhoz\ the
signal and background charge asymmetries are free parameters in the fit.

\begin{figure}[!bp]
\begin{center}
  \includegraphics[width=1.0\linewidth]{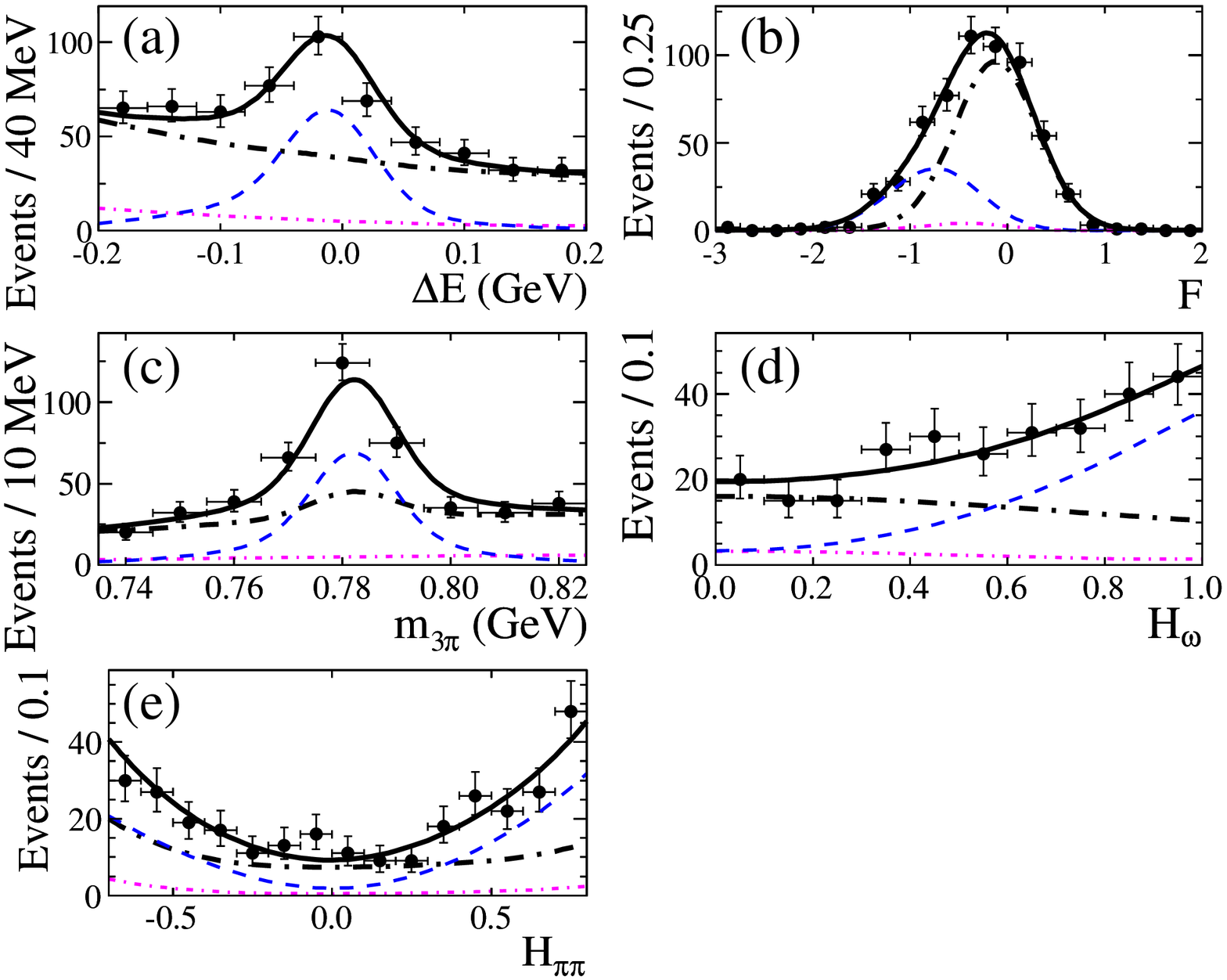}
\caption{Projections for \omegarhop: (a) \DE, (b) \xf, 
(c) $m_{3\pi}$, (d) \helo, and (e) \helrho.
Points with errors represent data and solid curves represent the full
fit functions. Also shown are signal (blue dashed), $b\to c$ background
(magenta dot-dashed), and total background (black long-dash-dotted).
Charmless background and nonresonant background are too small to be seen.
To suppress background, the plots are made with requirements on $\ln{\calL}$ 
that have an efficiency for signal of (40-60)\% depending on the plot.}
  \label{fig:proj_omegarhop}
\end{center}
\end{figure}
\begin{figure}[!htbp]
\begin{center}
  \includegraphics[width=1.0\linewidth]{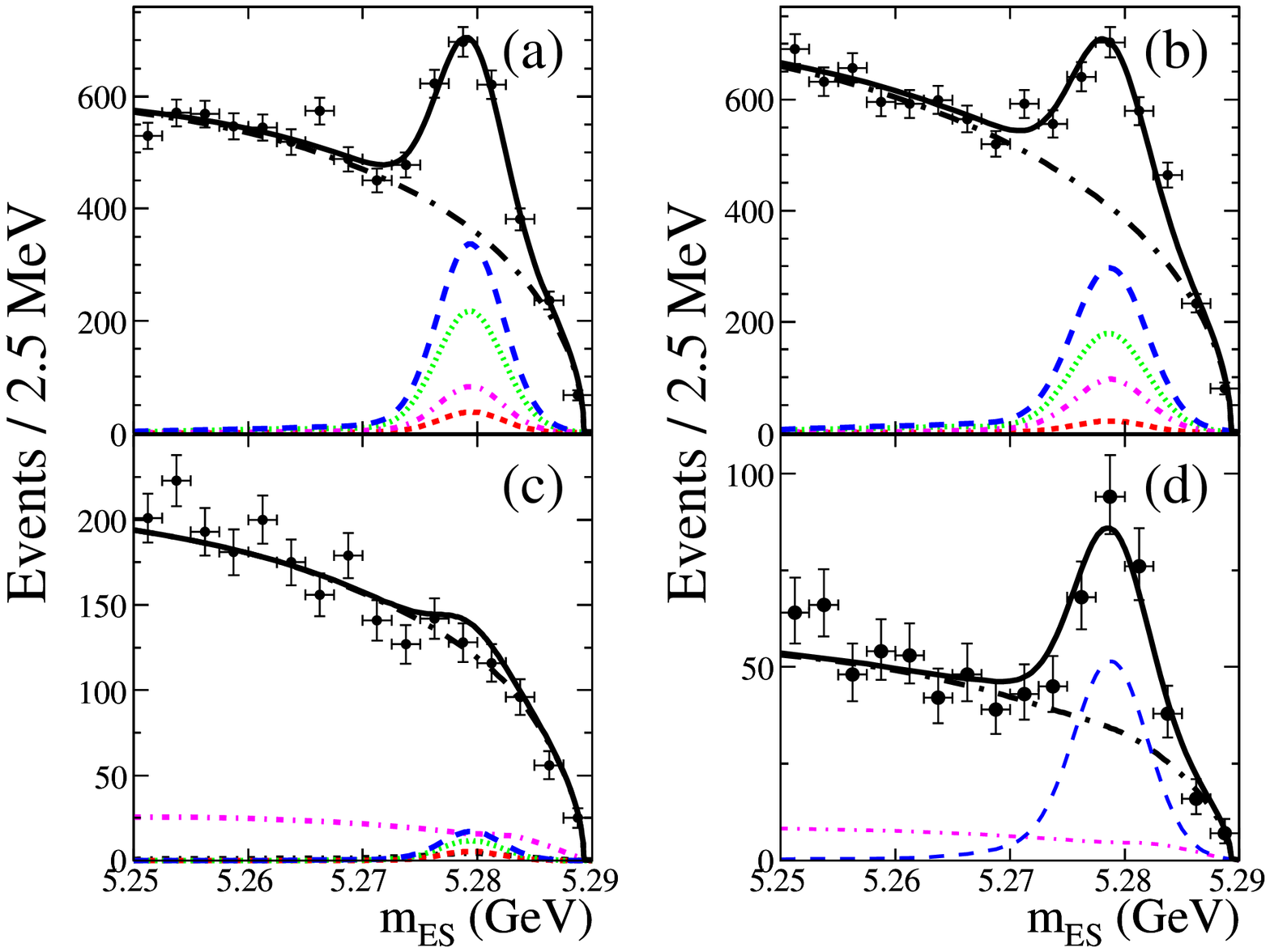}
  \caption{$B$-candidate \mes\ projections for (a) \fomegaKstz, (b) \fomegaKstp,
    (c) \fomegarhoz/\fomegafz, (d) \fomegarhop.
    The solid curve is the fit function, black long-dash-dotted is the
    total background, and the blue dashed curve is the total signal
    contribution.  For (a,b) we also show the signal components: \Kst(892)
    (red dashed), \KpiSwave\ (green dotted), and \Ktwost\ (magenta dot-dashed).
    We show for (c,d) the $b\to c$ background (magenta dot-dashed), and for (c)
    the \omegarhoz\ (red dashed) and \omegafz\ (green dotted) components.
    The plots are made with a requirement on $\ln{\calL}$ that has an 
    efficiency of (40-60)\% depending on the plot.}
  \label{fig:proj_mES}
\end{center}
\end{figure}
\begin{figure}[!htbp]
\begin{center}
  \includegraphics[width=1.0\linewidth]{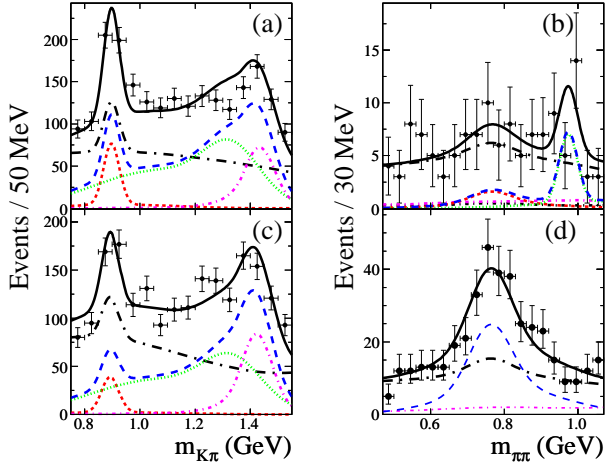}
  \caption{$B$-candidate $K\pi$ mass projections for (a) \fomegaKstz, 
   (c) \fomegaKstp, and $\pi\pi$ mass projections for (b) \fomegarhoz/\fomegafz,
   (d) \fomegarhop.  The efficiency range and description of the curves are 
   the same as for Fig.~\ref{fig:proj_mES}.}
  \label{fig:proj_mKpi}
\end{center}
\end{figure}

To describe the PDFs, we use simple functions such as the sum of two Gaussian
distributions for many signal components and the peaking parts of backgrounds,
low-order polynomials to describe most background shapes, an asymmetric
Gaussian for \xf, and the function
$x\sqrt{1-x^2}\exp{\left[-\xi(1-x^2)\right]}$ (with $x\equiv\mes/E_B^*$)
for the \mes\ background distributions.  These are illustrated for 
\omegarhop\ with projection plots of 
each fit variable in Figs.~\ref{fig:proj_omegarhop}, \ref{fig:proj_mES}d, and
\ref{fig:proj_mKpi}d.  
The parameters that determine the main features of the background PDF
shapes are allowed to vary in the fit.

We evaluate biases from our neglect of correlations among discriminating 
variables by fitting ensembles of simulated experiments.  Each
such experiment has the same number of events as the data for both background 
and signal; \qqbar\ background events are generated from their PDFs while 
signal and \BB\ background events are taken from fully simulated MC samples.
Since events from the \BB\ background samples are included in the ensembles, 
the bias includes the effect of these backgrounds.

We compute the branching fraction \calB\ for each decay by
subtracting the yield bias $Y_0$ from the measured yield, and dividing the
result by the efficiency and the number of produced \BB\ pairs.
We assume that the branching fractions of the \UfourS\ to \BpBm\ and \BzBzb\ 
are each equal to 50\%.
In Table~\ref{tab:results} we show for each decay mode the measured \calB, 
\fL, and \acp\ together with the quantities entering into these computations.  
For decays with \Kstarp\ we combine the results from the two \Kstar\ decay
channels, by adding their values of $-2\ln{\cal L}$.
For the significance $S$ we use the difference between the value of 
$-2\ln{\cal L}$
for zero signal and the value at its minimum; the corresponding
probability is interpreted with the number of degrees of freedom equal to two 
for modes with a measured  \fL\ and one for the others.  For modes without a 
significant signal, we quote a 90\%\ confidence level (C.L.) upper limit,
taken to be the branching fraction below which lies 90\% of the total
of the likelihood integral in the region of positive branching fraction.
In all of these calculations ${\cal L}(\calB)$ is a convolution of the
function obtained from the fitter with a Gaussian function
representing the correlated and uncorrelated systematic errors detailed
below. 

\begin{figure}[!htbp]
\begin{center}
  \includegraphics[width=1.0\linewidth]{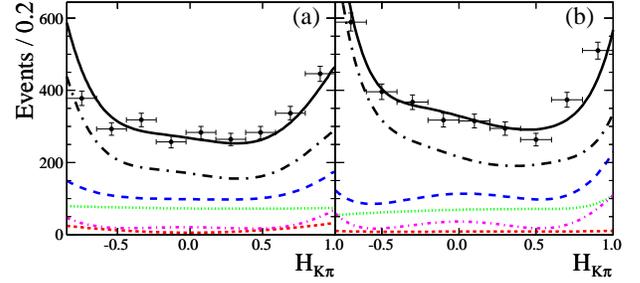}
  \caption{$B$-candidate $K\pi$ helicity projections for (a) \fomegaKstz, 
   and (b) \fomegaKstp.  The efficiency range and key for the curves are the 
   same as for Fig.~\ref{fig:proj_mES}(a,b).}
  \label{fig:proj_hKpi}
\end{center}
\end{figure}

We show in Fig.~\ref{fig:proj_mES} the data and PDFs projected onto \mes.
Figure \ref{fig:proj_mKpi} shows similar projections for the $K\pi$ and
$\pi\pi$ masses.  Figure \ref{fig:proj_hKpi}\ gives projections
onto \hel\ for the $\omega\Kst$ modes.


The systematic uncertainties on the branching fractions arising from
lack of knowledge of the signal PDF parameters are estimated 
by varying these parameters within uncertainties obtained from the 
consistency of fits to MC and data control samples.
The uncertainty in the yield bias correction is taken to
be the quadratic sum of two terms: half the bias correction and the
statistical uncertainty on the bias itself.  We estimate the uncertainty 
from the modeling of the nonresonant and \BB\ backgrounds by varying the 
background yields by their estimated uncertainties (from Ref.
\cite{PDG2008} and studies of our data).  We vary the SXF fraction by its 
uncertainty; we find this to be 10\% of its value, determined from studies of 
the control samples.  For the $\Kst_0(1430)$ modes, we vary the LASS parameters
within their measured uncertainties \cite{LASS}. For \omegarhoz\ 
where \fL\ is fixed, the uncertainty due to the
assumed value of \fL\ is evaluated as the change in branching fraction 
when \fL\ is varied by $^{+0.2}_{-0.3}$. 
These additive systematic errors are 
dominant for all modes and are typically similar in size except for the
error due to \BB\ background, which is usually smaller than the others.

Uncertainties in reconstruction efficiency, found from studies
of data control samples, are 0.4\%/track, 3.0\%/\piz, and 1.4\%/\KS\ decay. 
We estimate the
uncertainty in the number of $B$ mesons to be 1.1\%.  Published data
\cite{PDG2008}\ provide the uncertainties in the $B$-daughter
branching fractions ($\lsim$2\%).  The uncertainty in the efficiency of the 
\costhr\ requirement is (1.0--1.5)\%.
Since we do not account for interference among the \Kst\ components, we 
assign systematic uncertainties based on separate calculations where we vary 
the phases between the three components over their full range.

The systematic uncertainty on \fL\ includes the
effects of fit bias, PDF-parameter variation, and \BB\ and nonresonant
backgrounds, all estimated with the same method as used for the yield
uncertainties described above.  From large inclusive kaon and
$B$-decay samples, we estimate the \acp\ bias to be negligible for pions
and $-0.01$ for kaons, due primarily to material interactions.  Thus we correct 
the measured \acp\ for the \Kst\ modes by $+0.01$.  The
systematic uncertainty for \acp\ is estimated to be
$0.02$ due mainly to the uncertainty in this bias correction.  This
estimate is supported by the fact that the
corrected background \acp\ is smaller than 0.015.

In summary, we have searched for nine charmless hadronic $B$-meson decays
as shown in Table \ref{tab:results},
and have observed most of them (for the first time in all cases except
\omegarhop).  We calculate the branching fractions for $\omega\Kst_0(1430)$ 
using the composition of \KpiSwave\ from Ref.~\cite{Latham}.  We find 
$\BomegaKzstz=\RomegaKzstz$ and $\BomegaKzstp=\RomegaKzstp$, where the third 
errors arise from uncertainties in the branching fraction $K^*_0(1430)\ra K\pi$
\cite{PDG2008} and the resonant fraction of \KpiSwave.  For most decays
we measure \acp\ and find it to be consistent with zero.
For $VV$ and $VT$ decays we also measure \fL.   For \omegarhop,
\fL\ is near 1.0, as it is for $B\to\rho\rho$
\cite{rhorho}.  For the $VT$ \omegaKtst\ decays \fL\ is
about 4$\sigma$ from 1.0 for both charge states; it is similar to the value of
$\sim$0.5 found in $B\to\phi\Kst$ decays.
Branching fraction results are in agreement with
theoretical estimates \cite{BRYCYlilu} except for the $SV$ and $VT$
decays where the estimates are more uncertain \cite{KLO,CCY}.

We are grateful for the excellent luminosity and machine conditions
provided by our \pep2\ colleagues, 
and for the substantial dedicated effort from
the computing organizations that support \babar.
The collaborating institutions wish to thank 
SLAC for its support and kind hospitality. 
This work is supported by
DOE
and NSF (USA),
NSERC (Canada),
CEA and
CNRS-IN2P3
(France),
BMBF and DFG
(Germany),
INFN (Italy),
FOM (The Netherlands),
NFR (Norway),
MES (Russia),
MEC (Spain), and
STFC (United Kingdom). 
Individuals have received support from the
Marie Curie EIF (European Union) and
the A.~P.~Sloan Foundation.


\begin{thebibliography}{99}

\bibitem{phiKstorig}
\babar\ Collaboration, B. Aubert \etal, \jprl{91}, 171802 (2003);
Belle Collaboration, K.F.~Chen \etal, \jprl{91}, 201801 (2003).

\bibitem{BRYCYlilu}
M.~Beneke, J.~Rohrer, and D.~Yang, \npb{774}, 64 (2007);
H.-Y.~Cheng and K.-C.~Yang, \jprd{78}, 094001 (2008);
Y.~Li and C.-D.~L\"{u}, \jprd{73}, 014024 (2006).

\bibitem{rhorho}
Belle Collaboration, A. Somov \etal, \jprl{96}, 171801 (2006);
\babar\ Collaboration, B. Aubert \etal, \jprd{76}, 052007 (2007);
Belle Collaboration, J. Zhang \etal, \jprl{91}, 221801 (2003);
\babar\ Collaboration, B. Aubert \etal, \jprl{97}, 261801 (2006).

\bibitem{PrevBABAR}
\babar\ Collaboration, B. Aubert \etal, \jprd{74}, 051102 (2005).

\bibitem{VVBSMrefs}
C.W.~Bauer \etal, \jprd{70}, 054015 (2004);
P.~Colangelo, F.~De Fazio, and T.N.~Pham, \plb{597}, 291 (2004);
A.L.~Kagan, \plb{601}, 151 (2004);
M.~Ladisa \etal, \jprd{70}, 114025 (2004);
H.~Y.~Cheng, C.~K.~Chua, and A.~Soni, \jprd{71}, 014030 (2005);
H.-n.~Li and S.~Mishima, \jprd{71}, 054025 (2005);
H.-n.~Li, \plb{622}, 63 (2005).

\bibitem{nSMetc}
A.~K.~Giri and R.~Mohanta, \jprd{69}, 014008 (2004); 
E.~Alvarez \etal, \jprd{70}, 115014 (2004); 
P.~K.~Das and K.~C.~Yang, \jprd{71}, 094002 (2005);
C.-H.~Chen and C.-Q.~Geng, \jprd{71}, 115004 (2005); 
Y.-D.~Yang, R.~M.~Wang and G.~R.~Lu, \jprd{72}, 015009 (2005);
A.~K.~Giri and R.~Mohanta, \epjc{44}, 249 (2005);
S.~Baek \etal, \jprd{72}, 094008 (2005);
W.~Zou and Z. Xiao, \jprd{72}, 094026 (2005);
Q.~Chang, X.-Q.~Li, and Y.~D.~Yang, \jhep{0706}, 038 (2007).

\bibitem{oh}
S.~Oh, \jprd{60}, 034006 (1999).

\bibitem{BelleomKst}
Belle Collaboration, P. Goldenzweig \etal, \jprl{101}, 231801 (2008).

\bibitem{VVphaserefs}
D.~Atwood and A.~Soni, \jprd{59}, 013007 (1999);
D.~Atwood and A.~Soni, \jprd{65}, 073018 (2002);
H.-W.~Huang \etal, \jprd{73}, 014011 (2006).

\bibitem{BABARphiK2}
\babar\ Collaboration, B. Aubert \etal, \jprl{98}, 051801 (2007).

\bibitem{KLO}
C.~S.~Kim, J.-P. Lee, and S.~Oh, \jprd{67}, 014002 (2003).

\bibitem{etaKst}
\babar\ Collaboration, B. Aubert \etal, \jprl{89}, 201802 (2002).

\bibitem{CCY}
H.-Y.~Cheng, C.-K. Chua, and K.-C.~Yang, \jprd{77}, 014034 (2007).

\bibitem{Datta}
A.~Datta \etal, \jprd{77}, 114025 (2008).

\bibitem{BABARNIM}
\babar\ Collaboration, B.~Aubert \etal, \nima{479}, 1 (2002).

\bibitem{PDG2008}
Particle Data Group, C.~ Amsler \etal, \plb{667}, 1 (2008).

\bibitem{ccbarK0}
\babar\ Collaboration, B.\ Aubert \etal, \jprl{99}, 171803 (2007).

\bibitem{geant}
The \babar\ detector Monte Carlo simulation is based on GEANT4:
S.~Agostinelli \etal, \nima{506}, 250 (2003).

\bibitem{LASS}
LASS Collaboration, D.~Aston \etal, \npb{296}, 493 (1988).

\bibitem{Latham}
\babar\ Collaboration, B.\ Aubert \etal, \jprd{72}, 072003 (2005); 
{\bf 74}, 099903(E) (2006).

\bibitem{E791fz}
E791 Collaboration, E.~M. Aitala \etal, \jprl{86}, 765 (2001).


\end{thebibliography}
\end{document}